**RESEARCH ARTICLE**

**MATERIALS SCIENCE**

# Mass Production of Two-Dimensional Materials by Intermediate-Assisted Grinding Exfoliation

Chi Zhang[1+], Junyang Tan[1+], Yikun Pan[1], Xingke Cai[1], Xiaolong Zou[1], Hui-Ming Cheng[1,2,3]* and Bilu Liu[1]*

[1]Shenzhen Geim Graphene Center (SGC), Tsinghua-Berkeley Shenzhen Institute (TBSI) & Tsinghua Shenzhen International Graduate School (TSIGS), Tsinghua University, Shenzhen 518055, P. R. China.

[2]Shenyang National Laboratory for Materials Science, Institute of Metal Research, Chinese Academy of Sciences, Shenyang 110016, P. R. China.

[3]Advanced Technology Institute (ATI), University of Surrey, Guildford, Surrey GU2 7XH, UK.

[+]These authors contributed equally.

*E-mail: bilu.liu@sz.tsinghua.edu.cn, hmcheng@sz.tsinghua.edu.cn

## ABSTRACT

The scalable and high-efficiency production of two-dimensional (2D) materials is a prerequisite to their commercial use. Currently, only graphene and graphene oxide can be produced on a ton scale, and the inability to produce other 2D materials on such a large scale hinders their technological applications. Here we report a grinding exfoliation method that uses micro-particles as force intermediates to resolve applied compressive forces into a multitude of small shear forces, inducing the highly-efficient exfoliation of layer materials. The method, referred to as intermediate-assisted grinding exfoliation (iMAGE), can be used for the large-scale production of many 2D materials. As an example, we have exfoliated bulk h-BN into 2D h-BN with large flake sizes, high quality and structural integrity, with a high exfoliation yield of 67%, a high production rate of 0.3 g $h^{-1}$ and a low energy consumption of $3.01\times10^{6}$ J $g^{-1}$. The production rate and energy consumption are one to two orders of magnitude better than previous results. Besides h-BN, this iMAGE technology has been used to exfoliate various layer materials such as graphite, black phosphorus, transition metal dichalcogenides, and metal oxides, proving its universality. Molybdenite concentrate, a natural low-cost and abundant mineral, was used as a demo for the large-scale exfoliation production of 2D $MoS_2$ flakes. Our work indicates the huge potential of the iMAGE method to produce large amounts of various 2D materials, which paves the way for their commercial application.

## INTRODUCTION

The large number of two-dimensional (2D) materials, including graphene, hexagonal boron nitride (h-BN), transition metal dichalcogenides (TMDCs) like $MoS_2$ and $WSe_2$, metal oxides ($M_xO_y$), black phosphorene (b-P), *etc*, provide a wide range of properties and numerous potential applications,[1-7] but achieving these requires scalable production. Bottom-up strategies like chemical vapor deposition (CVD) and chemical synthesis have been extensively explored but only small quantities have been produced so far.[8,9] For example, the highest production of CVD-grown graphene is on the order of one million square meters (1,000,000 $m^2$) per year, which is equal to ~0.77 kg, while for other CVD-grown 2D materials it is much less.[10,11] Moreover, the high growth temperatures (450-1100 °C) and time-consuming transfer processes usually needed for CVD-grown 2D materials result in high cost and a low production rate.[11] As for the chemical synthesis strategy, solvothermal treatment has been used to synthesize different kinds of 2D materials including not only TMDCs[12] but also even hard-for-CVD-synthesised $M_xO_y$[13] and b-P[14], however, the quantity is still small. Another strategy for the preparation of 2D materials is top-down exfoliation, which involves exfoliating bulk layer materials to a monolayer or few-layer flakes.[3,7,15-18] Micromechanical exfoliation using the Scotch® tape can produce 2D materials with the highest quality, but in negligible amounts.[19] Although there are other techniques to exfoliate layer materials which can be scaled-up, they are only suitable for specific materials and/or require harsh conditions.[2,3,18,20] So far, only graphene and graphene oxide can be prepared at the ton level and the difficulties in preparing other 2D materials in such a large quantity has significantly hindered their commercialization.[11] Among the current exfoliation techniques, ball milling[17,21] and liquid phase exfoliation[22,23] which use shear forces to induce slipping between adjacent layers, are promising for the scalable production of 2D materials other than graphene, such as h-BN and $MoS_2$. However, ball milling usually needs a long time (e.g., 20 hours) to achieve high-yield (85%) exfoliation and the 2D materials obtained are small (~100 nm in lateral size) due to the large impact force of the irregularly moving hard balls[17] On the other hand, liquid phase exfoliation can produce 2D materials of high quality in a large scale,[22,24,25] but has a very low yield (< 3%). In addition, materials produced have low concentrations (<0.1 mg $mL^{-1}$)[24,25] in organic/surfactant solvents, and may need additional condensation and/or separation to concentrate them for use. It is obvious that both ball milling and liquid phase exfoliation still face many challenges for the scalable production of good quality 2D materials. An ideal exfoliation method for the commercial scalable production of 2D materials should meet the following requirements. First, the exfoliated products should have a large lateral size and high quality with good structural integrity, because size and quality are the two main parameters that influence the material's properties. Second, the method should be efficient, with a high exfoliation yield and a high production rate. Third, the process should be cheap, green and require a low energy input, in particular when considering large scale production. Fourth, it would be ideal if the method could be universal for the exfoliation of different layer materials. Unfortunately, such a method is still not available.

Note that layer materials like graphite and $MoS_2$ possess low coefficients of friction and have been used as solid lubricants in industry for more than half a century.[26] They can buffer a large compressive force ($F_c$) between rolls by releasing a frictional force ($F_f$), transferred from $F_c$, through slip between adjacent layers in their structure.[27] By increasing $F_c$, $F_f$ increases accordingly and eventually leads to the failure of the lubricant, which means irreversible sliding of the layers, i.e., exfoliation of these layer lubricants into thin 2D flakes. In principle, such a process would not have a big impact on the quality of the exfoliated material, because $F_f$ is parallel to planes of the material and no additional chemicals are involved during the process. Note that a large force is not needed to exfoliate layer materials because of the weak van der Waals interactions between their layers, as revealed by experimental measurements[28] and theoretical calculations[29]. For example, for a graphite flake with a lateral size of 10 μm, slip between layers happens at a small shear force of 3.9 μN[28] In comparison, the shear force provided by milling rolls is on the order of tens to hundreds of Newtons, over 6-8 orders of magnitude higher than the force needed to exfoliate graphite. Therefore, the critical factor in increasing the exfoliation efficiency is how to effectively produce a frictional force ($F_f$) on a layer material.

Based on the above analysis, we have developed an exfoliation technology which we call intermediate-assisted grinding exfoliation (iMAGE) that essentially meets these requirements. The key to our method is to use intermediate materials that increase the coefficient of friction of the mixture and effectively apply sliding frictional forces to the layer material, resulting in a dramatically increased exfoliation efficiency. Considering the case of 2D h-BN, the production rate and energy consumption can reach 0.3 g $h^{-1}$ and $3.01\times10^6$ J $g^{-1}$, respectively, both of which are one to two orders of magnitude better than previous results. The resulting exfoliated 2D h-BN flakes have an average thickness of 4 nm and an average lateral size of 1.2 μm. This iMAGE method has been extended to a series of layer materials with different properties, including graphite, b-P, TMDCs, and metal oxides. To further demonstrate this scaling up, molybdenite concentrate, a naturally existing cheap and earth abundant mineral, has been used for the large-scale exfoliation production of 2D $MoS_2$, making the commercial production and application of 2D $MoS_2$ feasible.

**RESULTS AND DISCUSSION**

Fig. 1a shows a schematic of the iMAGE method, and the experimental details are described in the Methods section. In short, the intermediates convert a macroscopic compressive force $F_c$ into microscopic forces $f_i$, on a layer material, so that $F_c = \sum_{i=1}^{n} f_i$, where $n$ is the number of microscopic forces in each box enclosed by the dashed line. The microscopic forces on each layer in the box must balance in the pressure direction, namely $F_c = \sum_{i=1}^{n} f_i = \sum_{i=1}^{n} f_i' = F_c'$, where $f_i'$ is a microscopic force from the balancing force of the bottom $F_c'$. Under grinding, the rotation of the platter at the bottom of the instrument induces slipping between the layer material and the force intermediate, and $f_i$ is converted into a sliding frictional force $f_{fi}$, where $f_{fi} = \mu f_i$, where $\mu$ is the sliding coefficient of friction between the intermediate and the layer material (Fig. 1b). When the shear friction force $f_{fi} > bE_e$, where $b$ and $E_e$ are the width and exfoliation energy of the layer material, the layers will slip and become exfoliated (Fig. S1), analogous to the failure of the solid lubricants discussed above.[28]

The effectiveness of the iMAGE method was evaluated by exfoliating h-BN because its exfoliation into a few layers (thickness < 5 nm) with a high yield and high quality is challenging.[17,20,21,30,31] In a typical experiment, bulk h-BN was ground with silicon carbide (SiC) particles, which were used as the force intermediate, using an apparatus that can provide a compressive force on the order of hundreds of Newtons with a rotation speed of 200 rpm (See Methods and Fig. S2). After grinding, the mixture contained exfoliated and un-exfoliated h-BN, and the SiC intermediate. To isolate the exfoliated 2D h-BN, the mixture was dispersed in deionized (DI) water, and after standing for 8 hours a green sediment containing SiC and un-exfoliated h-BN was seen, and the supernatant was milky white (Fig. S3). The observation of a clear Tyndall effect in the supernatant indicates the colloidal state of the exfoliated 2D h-BN in DI water (Fig. 1c). UV-vis-NIR optical absorption measurements of this supernatant showed an optical bandgap of 5.8 eV based on the plot of $(ah\nu)^2$ versus photon energy $h\nu$, indicating the dispersed material is high-quality h-BN with its original structure, which is confirmed by its white color (Figs. 1c and S3).[32] To verify the production of exfoliated 2D h-BN, SiC particles and bulk h-BN were independently added to DI water, and both precipitated completely within 5 minutes (Fig. S4) with a clear supernatant. These results confirm the separation of 2D h-BN from the mixture by simply dispersing it in water and standing. Previous studies show that nanomaterials can be dispersed in solvents when they have a surface energy that matches that of the solvent. The surface energy of 2D h-BN is in the range 44-66 mJ $m^{-2}$ [16] and that of DI Water is 72 mJ $m^{-2}$ so this is not the best match, but it is the most commonly used moderate solvent and has an acceptable solvability for 2D h-BN. In contrast, SiC and h-BN particles cannot be stabilized in DI water due to a mismatch of surface energy, and quickly precipitate because they have densities larger than the solvent, i.e., 3.2 g $cm^{-3}$ for SiC and 2.3 g $cm^{-3}$ for h-BN, while that for DI water is only 1.0 g $cm^{-3}$.

The lateral size, thickness, and quality of the 2D h-BN produced are important parameters for evaluating of this iMAGE technique (Figs. 1d and S5). Statistical analysis of the products by atomic force microscopy (AFM) indicates that the 2D h-BN has an average lateral size of 1.2 μm (Fig. 1e), consistent with the dynamic light scattering (DLS) results (Fig. S6), and an average thickness of 4 nm (Fig. 1f). To the best of our knowledge, this average lateral size is the largest so far obtained for top-down exfoliated 2D h-BN (Table S1). In addition, the results from high resolution transmission electron microscopy (HRTEM) observations and fast Fourier transformation (FFT) analysis show high quality 2D h-BN sheets (Fig. 1g) without noticeable defects in the planes and along the edges. The crystalline quality of the 2D h-BN is comparable with that of samples prepared by liquid phase exfoliation[22] and is superior to that of the samples obtained from intercalation[33] and ball milling[21], which have many structural defects and functional groups. The clear lattice fringes on the edge indicate negligible damage to the layer structure when transforming bulk h-BN into thin 2D sheets (Figs. 1h and S7). The large size and high quality of the obtained 2D h-BN must result from the unique exfoliation mode in the iMAGE process. During exfoliation, the SiC particles contact and interact with the top layers of bulk h-BN, causing them to slip with negligible damage to their in-plane structure. Fig. 1h shows a ten-layer h-BN step (region ii) slipping from twenty-five-layer h-BN (region iii), similar to previous *in-situ* TEM studies of the exfoliation of $MoS_2$, where the sliding of 2D sheets from the mother $MoS_2$ has been clearly identified under a shear force[34], consistent with the exfoliation mechanism proposed above.

Various spectroscopic characterization techniques confirm that the iMAGE-produced 2D h-BN has good quality with no noticeable functional groups (Figs. 1i-j; Supplementary Information, Section 1.1, Figs. S7 and S8). For example, X-ray photoemission spectroscopy (XPS) of 2D h-BN show two major peaks at 190.2 and 398.2 eV for B1s and N1s, the same as those for bulk h-BN, without any functional group-related peaks (Fig. 1i). Fourier transformation infrared (FTIR) analysis directly proves the absence of functional groups on the iMAGE-produced 2D h-BN (Fig. 1j). We note that the shapes of FTIR peaks for exfoliated 2D h-BN and bulk h-BN are different. This could be attributed to the reduced size and thickness after exfoliation. Compared to the lateral size of the exfoliated h-BN (1.2 μm), the size of the bulk h-BN (30 μm) is more close to the wavelength of infrared light in FTIR (0.2 μm ~ 20 μm), which would lead to a much stronger light scattering and cause the baseline drift as well as the peak broadening in bulk h-BN (Fig. 1j and Fig. S8). In addition, after exfoliation, due to the significantly reduced thickness, the B-N stretching and bending could be excited more easily, which will sharpen the FTIR peaks of the exfoliated h-BN.[35] Moreover, the Raman peak position of the B-N vibration mode ($E_{2g}$) for the exfoliated 2D h-BN is 1366 $cm^{-1}$, the same as that for bulk h-BN, also indicating negligible change to the intrinsic bonding (Fig. S7). These results are expected because iMAGE is a purely mechanical process in which no chemical reactions take place.

Based on the principle of friction force-induced sliding and exfoliation in the iMAGE technique, exfoliation occurs when $f_{fi} > bE_e$. Therefore, the parameters that influence $f_{fi}$, $b$ and $E_e$ will affect the exfoliation efficiency. It is clear that the size of the force intermediate will affect the exfoliation yield. For the bulk h-BN with a 30 μm particle size used in this study, the frictional force $f_{fi}$ needs to be larger than 13.8 μN assuming the exfoliation energy $E_e$ for h-BN and graphite are identical, i.e., 0.46 J $m^{-2}$.[28,29] We calculated the frictional force on h-BN applied by each SiC particle (Supplementary Information, Section 1.4 and Table S2), demonstrating that only particles larger than 73 μm can exert an average frictional force greater than 13.8 μN. This result shows that the sizes of the raw bulk materials and the intermediate particles should be matched to obtain a high exfoliation yield. An exfoliation yield of 67% is confirmed by both direct measurements of the weight of 2D h-BN and UV-vis absorption measurements (Supplementary Information, Section 1.2, Section 1.3, Fig. S9 and S10) which is one of the highest reported values for the exfoliation of pristine 2D h-BN (Supplementary Table S1), and it was achieved in 270 minutes. Such fast high-yield exfoliation can be understood when compared to the ball milling method (Supplementary Table S3). In the iMAGE process, the weight of the intermediates was almost an order of magnitude lower than that of the balls used in ball milling,[31] however, the number of particles is four orders of magnitude higher and the surface area is double, both of which facilitate the effective transfer of frictional force to the layer material, leading to a shorter exfoliation time with a much higher yield.

In addition to the exfoliation yield, production rate and energy consumption are the other two important parameters to be considered for the commercialization of the iMAGE method. Typically, when 2 g of bulk h-BN was used, the production rate of 2D h-BN could reach 0.3 g $h^{-1}$, which is around ten times that of a typical ball milling method for preparing amino-functionalized small 2D h-BN sheets[17] This rate is the highest value reported for h-BN production to date (Supplementary Table S1). The method achieves both a high yield and a high production rate (Fig. 2a), and this is the first time a 2D material other than graphene has been produced with a yield of more than 50% and a production rate of over 0.1g $h^{-1}$. An annual production capability of 2D h-BN is expected to be exceeding 10 tons by the iMAGE technology (Supplementary Information, Section 1.4). For comparison, electrochemical intercalation[18] and liquid phase exfoliation[22] can achieve only a high yield or a high production rate, while ball milling may produce functionalized 2D materials with a high yield but a medium production rate. In addition, our method requires an energy consumption of only $3.01 \times 10^6$ J $g^{-1}$ to exfoliate bulk h-BN, which is at least a tenth of that of other methods (Fig. 2b and Supplementary Table S1). According to theoretical calculations, the minimum energy required to exfoliate 1 g h-BN is 279 J (Supplementary Table S4), three orders of magnitude lower than the value for our method, demonstrating that there is still much room for improvement. In short, the iMAGE method requires a low energy to produce 2D materials because of its efficient transfer of force and energy to the layer materials, while the production rate is among the highest, making it a practical method for the commercial scalable production of 2D h-BN.

Besides SiC, several other hard micro-particles, such as $Mo_2C$ and cheap sea sand, were also used as intermediate materials to transmit the sliding fractional force onto bulk h-BN, and lead to the efficient exfoliation of h-BN, demonstrating that the economical production of 2D materials could be realized by using cheap intermediates. The process would also be valuable for the one-step fabrication of 2D material-based composites in which an intermediate is a functional component in the composite material.

In addition to h-BN, we have extended the method to a series of layer materials with a wide range of properties. Not only air-stable metallic graphite, *n*-type semiconducting $Bi_2Te_3$ and $MoS_2$, and insulating mica, but also air-sensitive *p*-type semiconducting b-P were exfoliated into few layers (Figs. 3a, S12, S13 and S14). UV-vis-NIR absorption (Fig. 3b) shows the wavelength-independent absorbance for the above materials in the visible light range. For graphene, the curve is similar to the Scotch® tape-exfoliated[36] and CVD-grown materials[37], but different from graphene oxide[4]. For as-exfoliated 2D $TiO_x$, mica and $Bi_2Te_3$, the characteristic absorption bands are almost at the same positions as those of the raw bulk materials (Supplementary Fig. S12) and previous reports[38,39] In addition, for $Bi_2Te_3$ the appearance of a prominent absorption peak at 265 nm reveals the production of few layer 2D $Bi_2Te_3$.[40] These results confirm that the iMAGE-produced 2D materials have good structural integrity rather than having been functionalized.

Regarding the size and thickness of the exfoliated 2D materials, taking $MoS_2$ as an example (Fig. S13), AFM measurements show that the exfoliated flakes have an average thickness of 10 nm and an average lateral size of 600 nm. The size of bulk raw $MoS_2$ used for this exfoliation was less than 2 μm, which is small. By using $MoS_2$ crystals with a larger size and optimizing the iMAGE process we could, in principle, obtain larger flakes. Raman analysis shows a redshift in the $A_{1g}$ phonon mode of 2D $MoS_2$ compared to bulk $MoS_2$, which is attributed to the lower interlayer van der Waals force after exfoliation [16]. The high quality of these exfoliated 2D $MoS_2$ is confirmed by HRTEM examination and UV-vis-NIR absorption with characteristic absorption peaks at 672 and 610 nm (Fig. S13).[41,42] The other layer materials (Figs. 3c-d and S12) were exfoliated into a few layers with an average thicknesses less than 10 nm and lateral sizes of 400-1000 nm under non-optimized conditions. Air-sensitive layer materials, such as b-P, have also been exfoliated by the iMAGE process in an argon- or nitrogen-filled glove box. The thickness of exfoliated 2D b-P is around 2-4 nm, as revealed by AFM characterization (Fig. 3d). UV-vis-NIR absorption spectrum shows a broad absorption peak around 520 nm (Fig. S14) as the wavelength range used has a much higher energy than the bandgap of b-P, which is similar to the previous reported results.[43] Raman spectroscopy shows that the $A_g^1/A_g^2$ ratios of exfoliated b-P and the bulk material were 0.88 and 0.68 (Fig. S14), respectively, which are both larger than the oxidation criterion value (0.6) and indicates that the 2D b-P flakes obtained were not oxidized after iMAGE exfoliation.[44]

As a step further, we used iMAGE to exfoliate abundant and cheap natural layer minerals to produce large quantities of 2D materials, for example $MoS_2$. The industrial production of $MoS_2$ ore follows the following procedure. 1) Finding a Mo-rich mine where the Mo concentration is higher than 0.1 wt% (Fig. 4a). 2) Drilling and blasting to obtain an ore with a size of a few meters (Fig. 4b). 3) Crushing the ore to a size of 10-20 μm. 4) Floating the ore to obtain a Mo-enriched molybdenite concentrate where Mo > 45 w% (Fig. 4c). This concentrate can then be used for the iMAGE exfoliation to prepare 2D $MoS_2$. Note that in the current molybdenum industry, the molybdenite concentrate is the flotation product of a natural ore and the precursor to industrial grade $MoS_2$. Because of its earlier production in the manufacturing processes, a molybdenite concentrate has a much lower price (~$\$2.6\times10^2$/t) than other molybdenum sources, e.g., 2.36% of the price of industrial-grade $MoS_2$ (~$\$1.1\times10^4$/t) and only 0.04% of the high-purity $MoS_2$ chemical from several vendors (~$\$7.0\times10^5$/t). The natural layer structure and low price suggest that molybdenite concentrate could be an ideal choice starting material for the scalable exfoliation of 2D $MoS_2$.

SEM inspection show that a bulk molybdenite concentrate does not have a well-defined layer structure because the main part is cubic particles with some layer-structured material on its surface (Fig. S15). By using the iMAGE method we prepared 60 liters of few-layer molybdenite dispersed in water (0.25 mg $mL^{-1}$) in less than 270 minutes (Fig. 4d). The concentrations of 2D materials could be further increased, for example, to be >5 mg $mL^{-1}$, by using suitable solvents or adding surfactants in water. AFM and TEM shows that the lateral size of the material was 200-300 nm (Fig. 4e). UV-vis-NIR absorption and Raman spectroscopy results confirmed the high quality of the $MoS_2$ exfoliated from a molybdenite concentrate (Figs. 4e and S15), and were the same as those obtained from $MoS_2$ exfoliated from $MoS_2$ chemicals (Figs. 3 and S13). These results show the scaling-up potential of the iMAGE method and suggest that it could be used to exfoliate cheap and abundant raw materials or natural minerals to produce high-quality 2D materials.

## CONCLUSION

We have developed a powerful iMAGE method to exfoliate various layer materials into 2D materials, which is suitable for commercial-scale production. The method converts applied macroscopic compressive forces into microscopic frictional forces with the assistance of an intermediate. These forces cause slipping of the layers in the layer materials resulting in highly-efficient exfoliation. Taking h-BN as an example, the iMAGE prepared 2D h-BN has a good quality, an average thickness of 4 nm (~12 layers), and average lateral size of 1.2 μm. Moreover, this method has one of the highest yields, the highest production rate, the lowest energy consumption, and the best scalable production ability of all currently-available techniques. The generality of this method has been proven by the effective exfoliation of several layer materials including graphite, $Bi_2Te_3$, b-P, $MoS_2$, $TiO_x$, h-BN, and mica, covering 2D metals, semiconductors with different bandgaps, and insulators. The scaling-up capability of the iMAGE technique has been demonstrated by exfoliating the cheap and abundant mineral, molybdenite, to produce 2D $MoS_2$ in large quantities. The iMAGE method overcomes one of the main challenges, scalable production, in the 2D materials field, and is expected to significantly accelerate their commercialization for a wide range of applications.

## METHODS

### Materials

h-BN (with an average size of 30 μm, Qinghuangdao ENO High-Tech Material Development Co. Ltd., China), graphite (325 mesh, Shanghai Macklin Biochemical Co. Ltd., China), mica (600 mesh, Shandong Usolf Chemical Tech. Co. Ltd., China), $MoS_2$ (<2 μm, Shanghai Macklin Biochemical Co. Ltd., China), *n*-type $Bi_2Te_3$ (1500 mesh, Wuhan Xinrong New Materials Co., Ltd, China), molybdenite concentrate (<15 μm, collected from the Sandaozhuang open-pit mine, Luoyang, China), SiC (150 mesh, purity >97%, Dongtai Mingzhi Silicon Carbide Co. Ltd., China) and sea sand (~1 mm, collected from a local beach in Shenzhen, China) were used as-received. The synthesis of bulk $TiO_x$ crystal is based on a previous work[45] with some modifications, and the details will be published later. In short, the preparation process could be divided into two stages, including (i) alkali metal ion intercalation and (ii) protonation of the titanates. In the stage (i), the layered $K_{0.8}Ti_{1.73}Li_{0.27}O_4$ was prepared by grinding and heating a stoichiometric mixture of $TiO_2$, $K_2CO_3$, and $Li_2CO_3$ at 1000 °C for two times. After the first calcination for 5 hours, the intermediate products need to be taken out and grinded, followed by the second calcination for 20 hours. In the stage (ii), the stage (i) obtained $K_{0.8}Ti_{1.73}Li_{0.27}O_4$ (1g) was added into HCl solution (1 M, 200 mL) and stirred for 4 days continuously. After stopping stirring, the product $H_{1.07}Ti_{1.73}O_4$ was obtained by washing the sediment with plenty of water and drying at room temperature. Then, the $H_{1.07}Ti_{1.73}O_4$ bulk underwent the iMAGE exfoliation process to prepare $TiO_x$ flakes. Bulk crystals of b-P were grown by mineralizer-assisted short-distance transport (SDT) reactions. Different liquids, including DI water (Millipore Milli-Q water purification system, 18.2 MΩ) and dimethylformamide (DMF, purity >99 wt%, Shanghai Macklin Biochemical Co. Ltd., China) were used without further purification.

**Detailed procedure of the iMAGE process**

In a typical process, raw h-BN particles (2 g) and force intermediates (SiC, 8 g) were added to a fast rotation apparatus (RM 200, Germany), and ground for 270 min under a compressive force of around 100 Newtons. To exfoliate air-stable layer materials like h-BN, graphite, $Bi_2Te_3$, $MoS_2$, $TiO_x$ and mica, the iMAGE process was performed in ambient atmosphere. To exfoliate air-sensitive layer materials, such as b-P, the iMAGE was performed in a glove box (500 × 600 × 600 $mm^3$), which was evacuated and then filled with argon or nitrogen gas twice before the exfoliation process. For large-scale exfoliation, a molybdenite concentrate (15 g, collected from the Sandaozhuang open-pit mine, Luoyang, China) and SiC (20 g) were ground together for 270 min under an ambient atmosphere.

The as-ground mixture was added to liquids (DMF for b-P and water for the others). After standing for 8 hours, a sediment was clearly seen at the bottom of the container, and the exfoliated 2D materials were stably dispersed in the supernatant. To collect the exfoliated 2D materials from the supernant, the dispersion was centrifuged at 1500 rpm for 30 minutes, and then vacuum filtered onto an alumina membrane with a pore size of 20 nm.

**Material characterization**

A UV-vis-NIR absorption spectrometer (Agilent Cary 5000, USA) was used to characterize the exfoliated 2D materials in dispersion. Optical microscopy (Carl Zeiss Axio Imager 2, Germany), atomic force microscopy (AFM, tapping mode, Oxford Instruments, USA), scanning electron microscopy (SEM, Hitachi SU8010, 15 kV, Japan), transmission electron microscopy (TEM, FEI Tecnai G2 F30, 300 kV, USA) and a laser particle size analyser (Malvern Zetasizer Nano-ZS90, UK) were used to characterize the morphology and structure of the exfoliated 2D materials, including their lateral size, thickness, crystal quality, etc. Raman spectroscopy (514 nm laser with a power density of 1 mW $cm^{-2}$, Horiba LabRAM ER, Japan) was used to examine the quality of the bulk and exfoliated 2D materials. X-ray photoelectron spectroscopy (XPS, monochromatic Al Kα X-rays, 1486.6 eV, PHI VersaProbe II, Japan) and Fourier transform infrared spectroscopy (FTIR, Thermo Scientific Nicolet iS 50, USA) were used to examine the chemistry of the bulk and exfoliated 2D h-BN. In addition, XPS and powder X-ray diffraction (XRD, with monochromatic Cu Kα radiation $\lambda = 0.15418$ nm, Bruker D8 Advance, Germany) were used to examine the purity and crystallinity of the exfoliated 2D materials collected from the supernatant.

## SUPPLEMENTARY DATA

Supplementary data are available at *NSR* online.

## ACKNOWLEDGEMENTS

We thank Mingqiang Liu for the growth of bulk b-P crystals by short-distance transport (SDT) method, and Dr. Peng Wang from the Luoyang Shenyu Molybdenum Co. Ltd. for providing raw $MoS_2$ and molybdenite concentrate from the Sandaozhuang open-pit mine. We also thank Profs. Feiyu Kang and Jingyu Xu for helpful discussions.

## FUNDING

This work was supported by the National Natural Science Foundation of China (Nos. 51722206, 51920105002, and 51521091), the Bureau of Industry and Information Technology of Shenzhen for the “2017 Graphene Manufacturing Innovation Center Project” (No. 201901171523), the Youth 1000-Talent Program of China, the Guangdong Innovative and Entrepreneurial Research Team Program (No. 2017ZT07C341), and the Development and Reform Commission of Shenzhen Municipality for the development of the “Low-Dimensional Materials and Devices” discipline.

## AUTHOR CONTRIBUTIONS

B.L. and H.M.C. conceived the idea and supervised the project. C.Z., J.T., Y.P. performed most materials exfoliation and characterization experiments. C.Z., J.T., X.C., X.Z., B.L., and H.M.C. analysed the results. C.Z., J.T., X.C., B.L., and H.M.C. wrote the manuscript with feedback from the other authors.

## REFERENCES

1. Cai, X., Luo, Y., Liu, B. & Cheng, H. M. Preparation of 2D material dispersions and their applications. *Chem. Soc. Rev.* **47**, 6224-6266 (2018).
2. Anasori, B., Lukatskaya, M. R. & Gogotsi, Y. 2D metal carbides and nitrides (MXenes) for energy storage. *Nat. Rev. Mater.* **2**, 16098 (2017).
3. Zhu, Y. *et al.* Carbon-based supercapacitors produced by activation of graphene. *Science* **332**, 1537-1541 (2011).
4. Li, D., Muller, M. B., Gilje, S., Kaner, R. B. & Wallace, G. G. Processable aqueous dispersions of graphene nanosheets. *Nat. Nanotech.* **3**, 101-105 (2008).
5. Zhu, Y., Ji, H., Cheng, H. M. & Ruoff, R. S. Mass production and industrial applications of graphene materials. *Natl. Sci. Rev.*, **5**, 90-101 (2017).
6. Watts M. C. *et al.* Production of phosphorene nanoribbons. *Nature*, **568**, 216-220 (2019).
7. Mounet, N. *et al.* Two-dimensional materials from high-throughput computational exfoliation of experimentally known compounds. *Nat. Nanotech.* **13**, 246-252 (2018).
8. Wu, T. *et al.* Fast growth of inch-sized single-crystalline graphene from a controlled single nucleus on Cu-Ni alloys. *Nat. Mater.* **15**, 43 (2016).
9. Fan, Z. *et al.* Surface modification-induced phase transformation of hexagonal close-packed gold square sheets. *Nat. Commun.* **6**, 6571 (2015).
10. Wang, X.Y., Narita, A. & Muellen, K. Precision synthesis versus bulk-scale fabrication of graphenes. *Nat. Rev. Chem.* **2**, 0100 (2017).

11. Ren, W. & Cheng, H. M. The global growth of graphene. *Nat. Nanotech.* **9**, 726-730 (2014).

12. Luo, Y. *et al.* Morphology and surface chemistry engineering toward pH-universal catalysts for hydrogen evolution at high current density. *Nat. Commun.* **10**, 269 (2019).

13. Sun, Z. *et al.* Generalized self-assembly of scalable two-dimensional transition metal oxide nanosheets. *Nat. Commun.* **5**, 3813 (2014).

14. Tian, B. *et al.* Facile bottom-up synthesis of partially oxidized black phosphorus nanosheets as metal-free photocatalyst for hydrogen evolution. *Proc. Natl. Acad. Sci. USA*, **115**, 4345 (2018).

15. Eda, G. & Chhowalla, M. Chemically derived graphene oxide: towards large-area thin-film electronics and optoelectronics. *Adv. Mater.* **22**, 2392-2415 (2010).

16. Coleman, J. N. *et al.* Two-dimensional nanosheets produced by liquid exfoliation of layered materials. *Science* **331**, 568-571 (2011).

17. Lei, W. *et al.* Boron nitride colloidal solutions, ultralight aerogels and freestanding membranes through one-step exfoliation and functionalization. *Nat. Commun.* **6**, 8849 (2015).

18. Zeng, Z. *et al.* An effective method for the fabrication of few-layer-thick inorganic nanosheets. *Angew. Chem. Int. Ed.* **51**, 9052-9056 (2012).

19. Novoselov, K. S. *et al.* Electric field effect in atomically thin carbon films. *Science* **306**, 666-669 (2004).

20. Zhu, W. *et al.* Controlled gas exfoliation of boron nitride into few-layered nanosheets. *Angew. Chem. Int. Ed.* **55**, 10766-10770 (2016).

21. Chen, S. *et al.* Simultaneous Production and Functionalization of Boron Nitride Nanosheets by Sugar-Assisted Mechanochemical Exfoliation. *Adv. Mater.*, **31**, 1804810 (2019).

22. Paton, K. R. *et al.* Scalable production of large quantities of defect-free few-layer graphene by shear exfoliation in liquids. *Nat. Mater.* **13**, 624-630 (2014).

23. Blake, P. *et al.* Graphene-based liquid crystal device. *Nano Lett.* **8**, 1704-1708 (2008).

24. Varrla, E. *et al.* Turbulence-assisted shear exfoliation of graphene using household detergent and a kitchen blender. *Nanoscale* **6**, 11810 (2014).

25. Bonaccorso, F., Bartolotta, A., Coleman, J. N. & Backes, C. 2D-crystal-based functional inks. *Adv. Mater.* **28**, 6136-6166 (2016).

26. Scharf, T. W. & Prasad, S. V. Solid lubricants: a review. *J Mater Sci* **48**, 511-531 (2013).

27. Prasad, S. V. & Zabinski, J. S. Tribology of tungsten disulphide ($WS_2$): characterization of wear-induced transfer films. *J. Mater. Sci. Lett.* **12**, 1413-1415 (1993).

28. Wang, W. *et al.* Measurement of the cleavage energy of graphite. *Nat. Commun.* **6**, 7853 (2015).

29. Björkman, T., Gulans, A., Krasheninnikov, A. V. & Nieminen, R. M. van der Waals bonding in layered compounds from advanced density-functional first-principles calculations. *Phys. Rev. Lett.* **108**, 235502 (2012).

30. Li, X. *et al.* Exfoliation of hexagonal boron nitride by molten hydroxides. *Adv. Mater.* **25**, 2200-2204 (2013).

31. Lee, D. *et al.* Scalable exfoliation process for highly soluble boron nitride nanoplatelets by hydroxide-assisted ball milling. *Nano Lett.* **15**, 1238-1244 (2015).

32. Weng, Q. *et al.* Tuning of the optical, electronic, and magnetic properties of boron nitride nanosheets with oxygen doping and functionalization. *Adv. Mater.* **29** (2017).

33. Voiry, D., Mohite, A. & Chhowalla, M. Phase engineering of transition metal dichalcogenides. *Chem. Soc. Rev.* **44**, 2702-2712 (2015).

34. Tang, D. M. *et al.* Nanomechanical cleavage of molybdenum disulphide atomic layers. *Nat. Commun.* **5**, 3631 (2014).

35. Du, M. *et al.* A facile chemical exfoliation method to obtain large size boron nitride nanosheets. *CrystEngComm,* **15**, 1782 (2013).

36. Nair, R. R. *et al.* Fine structure constant defines visual transparency of graphene. *Science* **320**, 1308 (2008).

37. Kim, K. S. *et al.* Large-scale pattern growth of graphene films for stretchable transparent electrodes. *Nature* **457**, 706-710 (2009).

38. Osada, M. *et al.* Controlled doping of semiconducting titania nanosheets for tailored spinelectronic materials. *Nanoscale* **6**, 14227-14236 (2014).

39. Pan, X. F. *et al.* Transforming ground mica into high-performance biomimetic polymeric mica film. *Nat. Commun.* **9**, 2974 (2018).

40. Srivastava, P. & Singh, K. Low temperature reduction route to synthesise bismuth telluride ($Bi_2Te_3$) nanoparticles: structural and optical studies. *J. Exp. Nanosci.* **9**, 1064-1074 (2014).

41. Zhao, W. *et al.* Lattice dynamics in mono- and few-layer sheets of $WS_2$ and $WSe_2$. *Nanoscale* **5**, 9677-9683 (2013).

42. Backes, C. *et al.* Edge and confinement effects allow in situ measurement of size and thickness of liquid-exfoliated nanosheets. *Nat. Commun.* **5**, 4576 (2014).

43. Woomer, A. H. *et al.* Phosphorene: synthesis, scale-up, and quantitative optical spectroscopy. *ACS Nano.* **9**, 8869 (2015).

44. Hanlon, D. *et al.* Liquid exfoliation of solvent-stabilized few-layer black phosphorus for applications beyond electronics. *Nat. Commun.* **6**, 8563 (2015).

45. Sasaki, T. *et al*. A mixed alkali metal titanate with the lepidocrocite-like layered structure. Preparation, crystal structure, protonic form, and acid-base intercalation properties. *Chem. Mater.* **10**, 4123-4128 (1998).

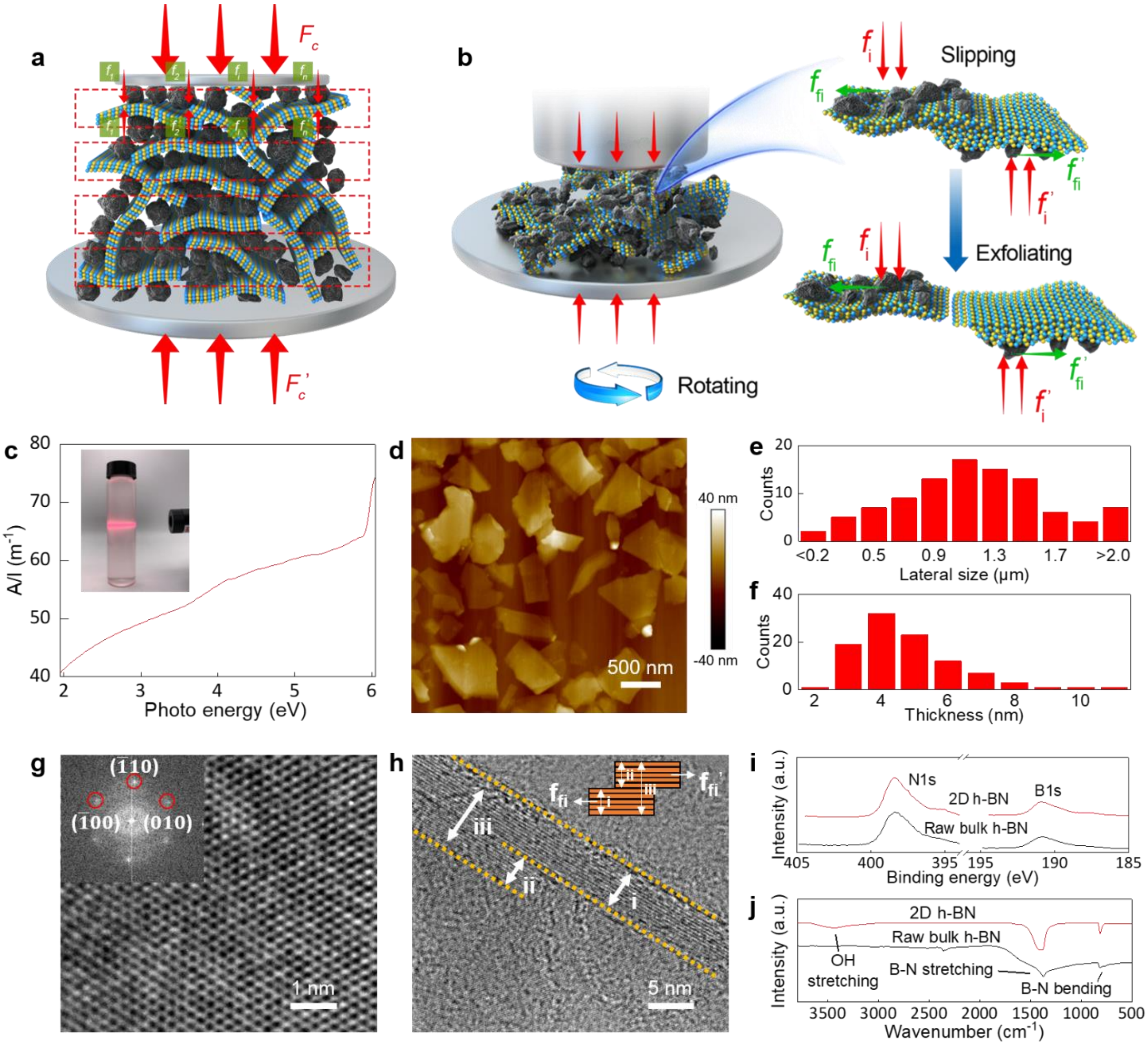

**Figure 1 | Exfoliation mechanism of the iMAGE method and characterization of as-prepared 2D h-BN. a,** Schematic of the decomposition of the macroscopic compressive forces $F_c$ and $F_c'$ into much smaller microscopic forces $f_i$ and $f_i'$ that were loaded onto the layer materials by force intermediates. **b,** Exfoliation mechanism of layer materials. $f_i$ and $f_i'$ transfer to sliding frictional forces $f_{fi}$ and $f_{fi}'$ under the relative slipping of the intermediates and layer materials due to the rotation of the bottom container. When $f_{fi} > bE_e$, exfoliation of the layer material occurs. Note here $b$ is the width of the layered materials, and the unit of $bE_e$ is (m)*(J*m$^{-2}$) = N. **c,** A UV-vis-NIR absorption spectrum of exfoliated 2D h-BN dispersed in DI water. Inset shows the Tyndall effect of the dispersion. **d-f,** AFM image, statistical analysis of the lateral size and thickness of 2D h-BN. **g,** A HRTEM image of the in-plane structure of exfoliated 2D h-BN and its FFT pattern (inset). **h,** Edge structure of the exfoliated 2D h-BN, showing that the region ii separates from region iii. Inset in **h** is an illustration

of the sliding and exfoliation process. **i-j,** XPS and FTIR spectra of the raw h-BN and exfoliated 2D h-BN.

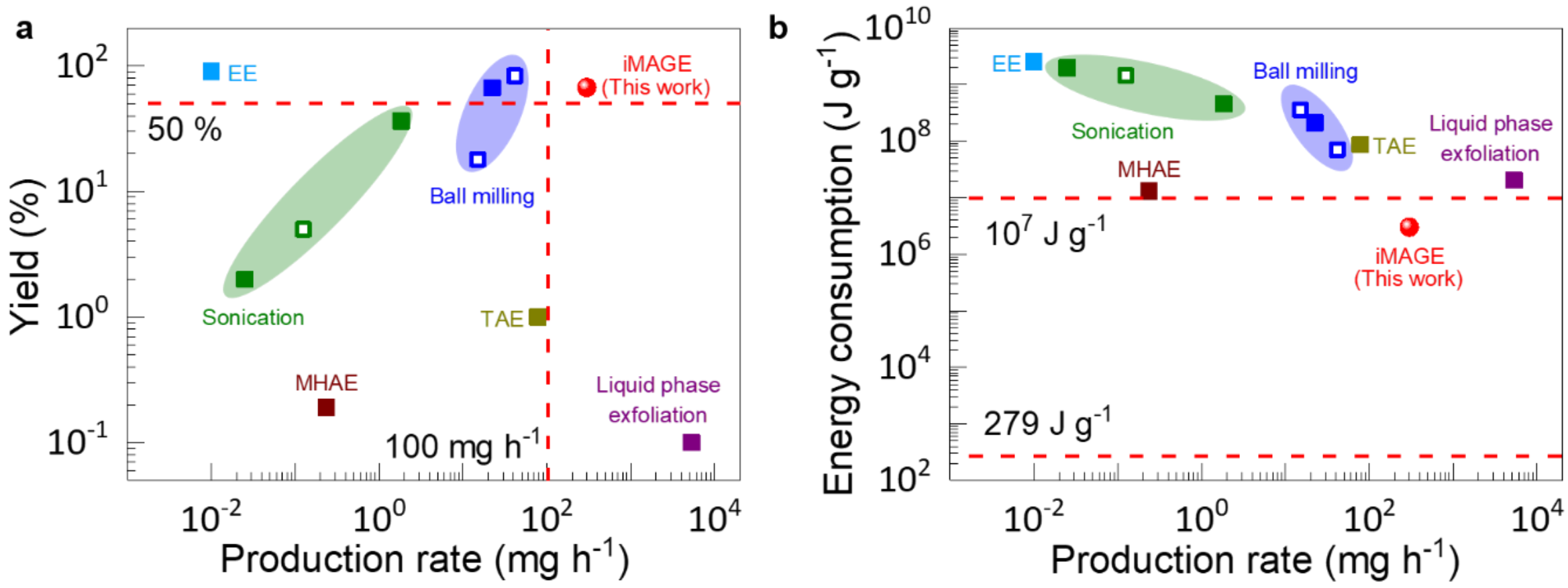

**Figure 2 | High-efficiency exfoliation of 2D h-BN by the iMAGE technique. a,** Exfoliation yield versus production rate and **b,** Energy consumption versus production rate, for different exfoliation methods, including ball milling, sonication, liquid phase exfoliation, electrochemical exfoliation (EE), turbulence-assisted exfoliation (TAE), molten hydroxide-assisted exfoliation (MHAE), and this method (iMAGE). Note that the solid (or empty) boxes stands for non-functionalized (or functionalized) 2D h-BN materials.

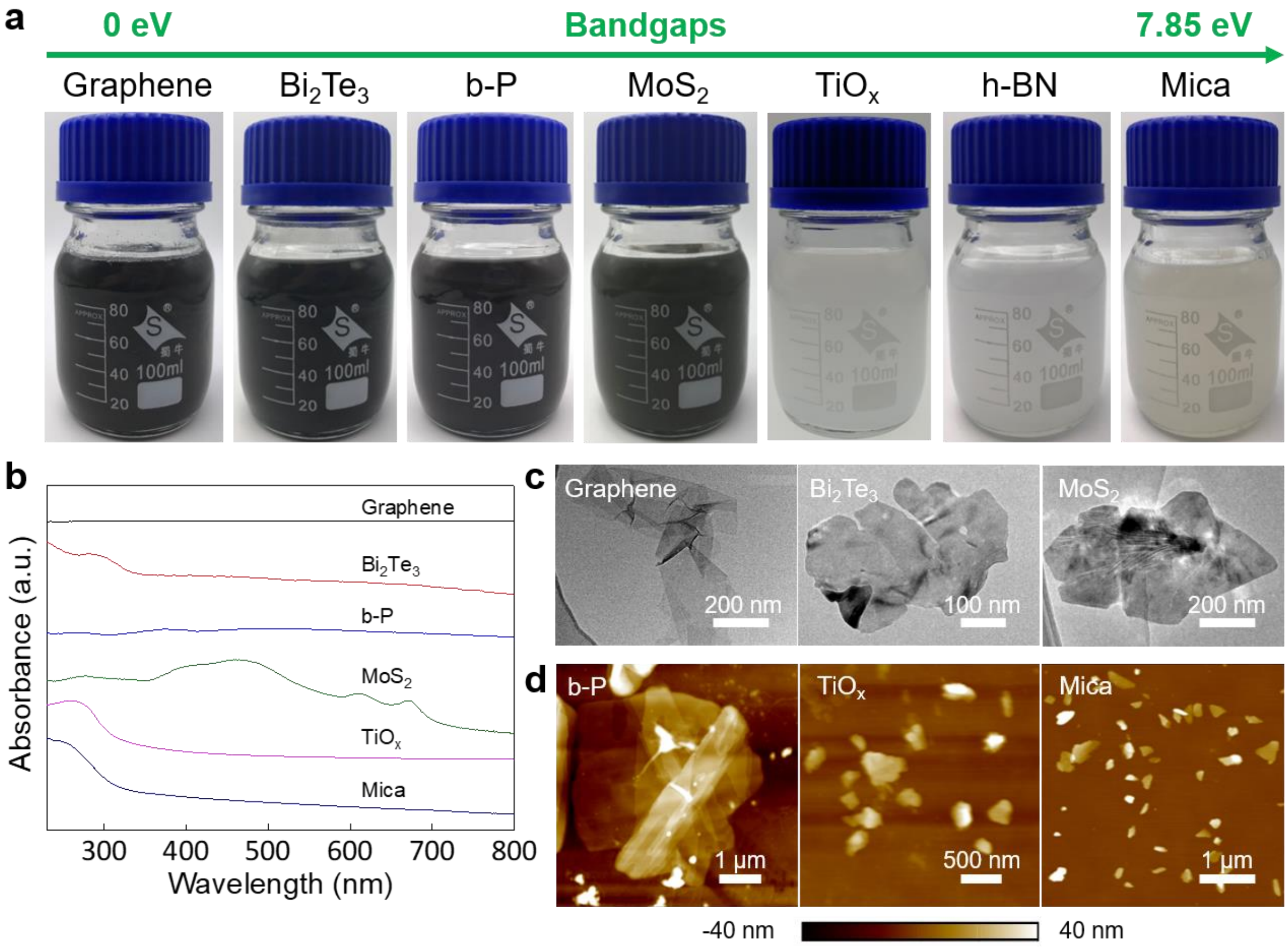

**Figure 3 | Extension of the iMAGE technique to exfoliate other layer materials. a,** Digital images, **b**, UV-vis-NIR absorption spectra, and **c-d,** TEM and AFM images of as-prepared 2D materials including metallic graphene, semiconducting $Bi_2Te_3$, b-P, $MoS_2$ and $TiO_x$, as well as insulating mica, confirming their 2D features.

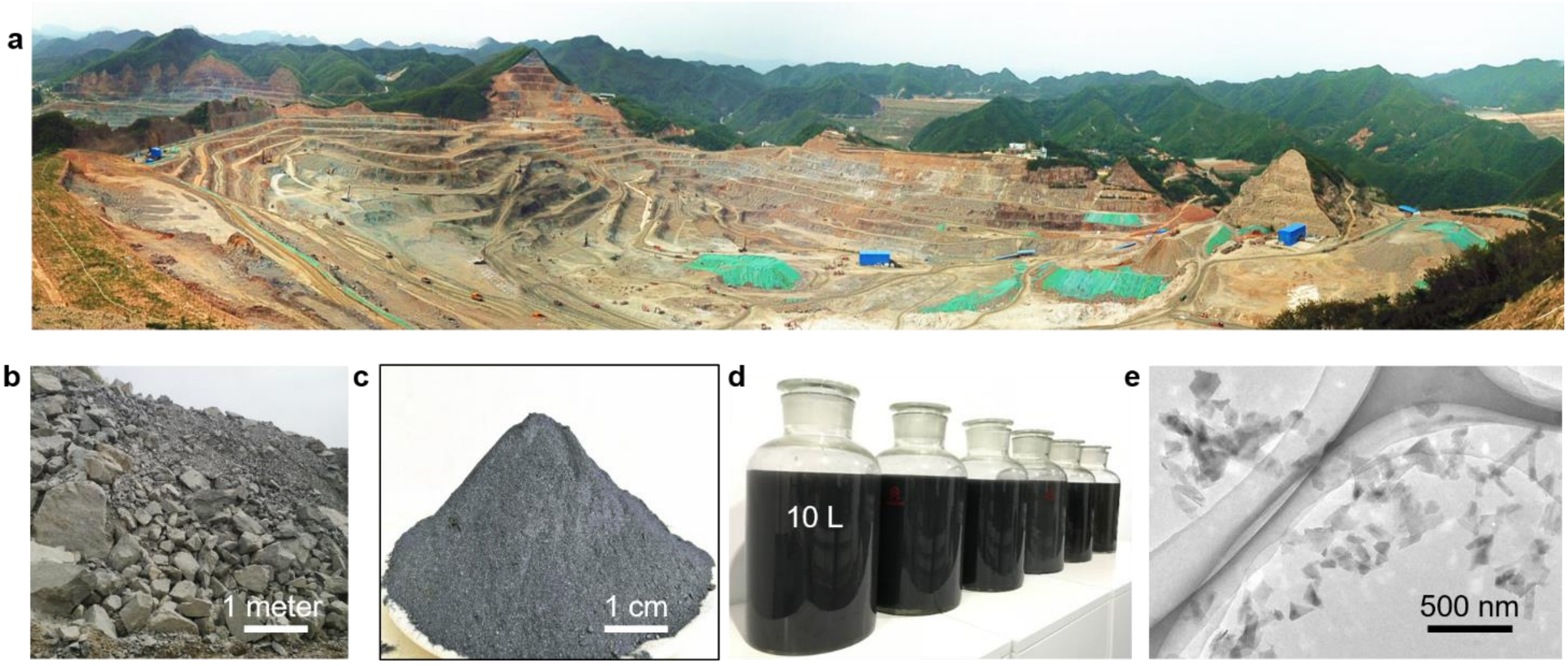

**Figure 4 | Large-scale production of 2D $MoS_2$ from a cheap molybdenite concentrate. a,** Bird's eye view of the Sandaozhuang open-pit mine in Luoyang, China. The size of the mine is around 1.5 kilometers by 1.5 kilometers, and the depth is around 400 meters. **b-d,** Digital images of ores, molybdenite concentrates, and 60 L of as-prepared 2D $MoS_2$ aqueous dispersions with a concentration of 0.25 mg $ml^{-1}$. **e,** TEM image of the exfoliated 2D $MoS_2$.